\documentclass[conference]{IEEEtran}
\IEEEoverridecommandlockouts

\makeatletter
\let\old@ps@headings\ps@headings
\let\old@ps@IEEEtitlepagestyle\ps@IEEEtitlepagestyle
\def\confheader#1{%
	\def\ps@headings{%
		\old@ps@headings%
		\def\@oddhead{\strut\hfill#1\hfill\strut}%
		\def\@evenhead{\strut\hfill#1\hfill\strut}%
	}%
	\def\ps@IEEEtitlepagestyle{%
		\old@ps@IEEEtitlepagestyle%
		\def\@oddhead{\strut\hfill#1\hfill\strut}%
		\def\@evenhead{\strut\hfill#1\hfill\strut}%
	}%
	\ps@headings%
}
\makeatother

\usepackage[pscoord]{eso-pic}
\newcommand{\placetextbox}[3]{
	\setbox0=\hbox{#3}
	\AddToShipoutPictureFG*{ \put(\LenToUnit{#1\paperwidth},\LenToUnit{#2\paperheight}){\vtop{{\null}\makebox[0pt][c]{#3}}}
	}
}
\placetextbox{.23}{0.055}{\small{To appear in Proc. ICEE2020}}

\usepackage{cite}
\usepackage{amsmath,amssymb,amsfonts}
\usepackage{algorithmic}
\usepackage{graphicx}
\usepackage{textcomp}
\usepackage{xcolor}
\usepackage{epstopdf}
\usepackage{balance}
\usepackage[left=1.57cm,right=1.57cm,top=1.9cm,bottom=2.54cm]{geometry}
\def\BibTeX{{\rm B\kern-.05em{\sc i\kern-.025em b}\kern-.08em
		T\kern-.1667em\lower.7ex\hbox{E}\kern-.125emX}}
\begin{document}
	
	\title{A Low Complexity Space-Time Block Codes  Detection for Cell-Free Massive MIMO Systems\\
	}
	
	
	\author{\IEEEauthorblockN{ Akbar Mazhari Saray\IEEEauthorrefmark{1},
			Jafar Pourrostam\IEEEauthorrefmark{1},
			 Seyed Hosein Mousavi\IEEEauthorrefmark{1},
			Mahmood Mohassel Feghhi\IEEEauthorrefmark{1}
}
		\IEEEauthorblockA{ Faculty of Electrical and Computer Engineering
			\\
			University of Tabriz\\
			Tabriz, Iran\\
			Email: \IEEEauthorrefmark{1}a.mazharisaray96@ms.tabrizu.ac.ir
			\\
			\IEEEauthorrefmark{1}j.pourrostam@tabrizu.ac.ir\\
			\IEEEauthorrefmark{1}mir.hosein.mousavi@tabrizu.ac.ir \\
			\IEEEauthorrefmark{1}mohasselfeghhi@tabrizu.ac.ir}}
	\maketitle
	
\begin{abstract}
	The new generation of telecommunication systems must provide acceptable data rates and spectral efficiency for new applications. Recently massive MIMO has been introduced as a key technique for the new generation of telecommunication systems. Cell-free massive MIMO system is not segmented into cells. Each BS antennas are distributed throughout the  environment and each user is served by all BSs, simultaneously.
	In this paper, the performance of the multiuser cell-free massive MIMO-system exploying space-time block codes in the uplink, and with linear decoders is studied. An Inverse matrix approximation using  Neumann series is proposed to reduce the computational and hardware complexity of the decoding in the receiver. 
	For this purpose, each user has two antennas, and also for improving the diversity gain performance, space-time block codes are used in the uplink. Then, Neumann series is used to approximate the inverse matrix in ZF and MMSE decoders, and its performance is evaluated in terms of BER and spectral efficiency.
	In addition, we derive lower bound for throughput of ZF decoder.
	The simulation results show that performance of the system , in terms of BER and spectral efficiency, is better than the single-antenna users at the same system. Also, the BER performance in a given system with the proposed method will be close to the exact method.
\end{abstract}


\section{INTRODUCTION}
Massive MIMO system is a key technique used in the 5G of the telecommunications networks
\cite{m1}.
It uses massive antenna arrays. By increasing the number of antennas and exploiting spatial diversity and beamforming, increases the spectral and energy efficiency of the system with simple signal processing \cite{m_aeu}. 
 It is one solution to compress a network. Cellular massive MIMO system performance is restricted by the intercellular interference that arises from the cellular structure  of this system.
A cell-free multi-user massive MIMO system has recently been proposed as an alternative to cell-based systems. In this way we will not have the standard cellularization for wireless communication
\cite{m2}.
In this system, base stations (BSs) are distributed all over the environment, and all of them are connected to a central controller unit(CPU) by a back-haul link providing unlimited capacity.
BSs are used to cover all users and only user data and power control coefficients is exchanged between BSs and CPU
\cite{m3 , m4 , m5}.

In point-to-point MIMO systems, the link capacity is proportional $ min(M,K) $, where $ M $ and $ K $ are the number of BS and user terminal antennas, respectively  \cite{m6 , m7}. As a result, single-antenna users will significantly reduce the actual capacity of the system, Therefore, equipping users with two antennas is suggested as a solution for improving the spectral efficiency of the system
\cite{m8}.

Dual-antenna users, provide two  channels between user and BS and if the number of BS antennas is high, then this channels will be independent. Because each user has two antennas, we can use space-time block code as a way to increase  the diversity gain in uplink, which for example in
\cite{m9 , m11 , m12, m_ijcs}
for  MIMO system have been investigated
\cite{m8}.

On the other hand, in designing, the target algorithms, implementation feasibility and hardware limitations are usually carefully evaluated. Therefore, one of the important issues in design is the application of low computational complexity methods with quasi-optimal performance. It should be noted that most of the computational complexity of the linear decoding in the receiver is on the inverse Hermitian matrix. Therefore, assuming the service is provided to many users using inverse matrix computation algorithms, such as Cholesky Decomposition, the system will have high computational complexity. Therefore, it is recommended to use Neumann series for efficient and fast calculation of approximate inverse matrix
\cite{m13}.

The 
\cite{m8}
shows that if the transmitted signals of each user are programmed correctly (ّ by choosing the space-time block code) and the number of each BS antennas is large, then the interference between users and two antennas of each user with linear decoders will be eliminated.

In this paper, we study the achievable  spectral efficiency and bit error rate (BER) of a multiuser cell-free massive MIMO system using space-time block codes with  linear decoders in the uplink. 
Finally, the approximate algorithm for matrix computation based on Neumann series is used to decrease the computational complexity of the receiver and the simulation results are evaluated and compared.
So, the paper structure  is as: section 
\ref{f}
introduces the system model, and section 
\ref{ff}
illustrates the design of space-time block codes, linear decoders and approximate calculation of the inverse decoding matrix. Section 
\ref{fff}
discusses the detailed results of the BER and achievable spectral   efficiency. The numerical results is presented in Section 
\ref{ffff}
and finally, section
\ref{dd} concludes the paper.

Notation: Vectors and matrices are written in Bold lowercase letters and boldface with capital letters, respectively.
The  Hermitian is presented by
the superscripts H. $ \mathbf{I}_K $ represents the $ K \times K $
identity matrix. The notation $ ||.|| $ and  $ \mathbb{E} {(.)} $ are used to represent Euclidean norm and the expectation of a random variable, respectively. The $ \mathbf{o} \sim \mathcal{CN} (0, C) $  is complex Gaussian vector with zero-mean and variance C.

\section{SYSTEM MODEL}
\label{f}
We consider  a multiuser cell-free massive MIMO system, in which the CPU communicates with all the BSs through a back-haul link, and only user data and power control coefficients is exchanged between them.  It is assumed that the back-haul link between the BSs and the CPU is an error-free link with unlimited capacity.
\begin{figure}[t]
	\centerline{\includegraphics[scale=.24]{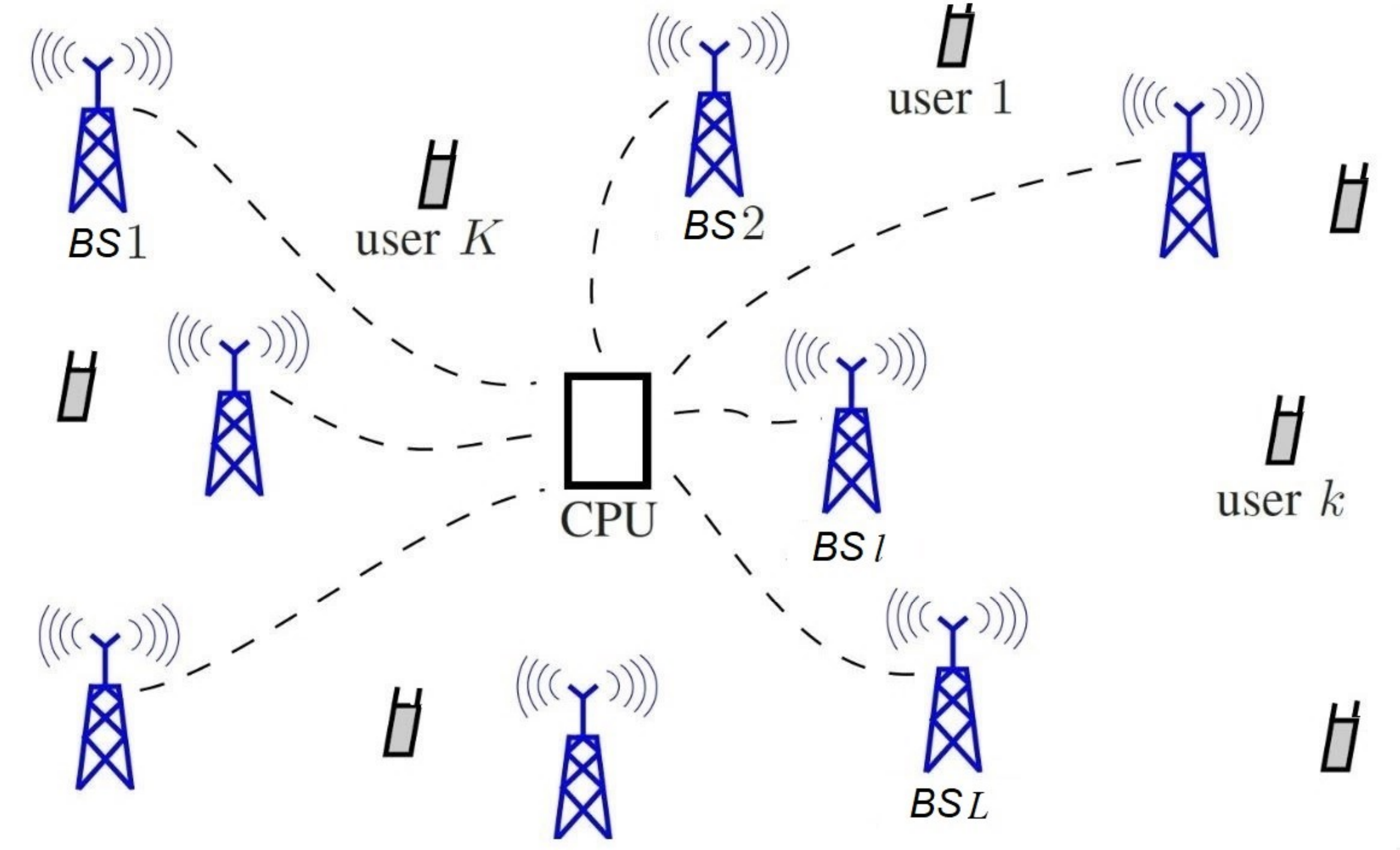}}
	\caption{Multiuser cell-free multiuser massive MIMO system}
	\label{fig1}
\end{figure}
The channel coefficients from the $ m $-th antenna in $ l $-th BS to the $ j $-th antenna in $ k $-th user is 
$ g_{mljk} = \beta_{mljk} h_{mljk} $, 
$ ( 1 \le k \le K , j=1,2 ,  1 \le m \le M  )$.
 In this system $  h_{mljk} $
 denotes the small scale fading that is assumed i.i.d. and is specified as 
 $\mathbf{h}_{mljk} \sim \mathcal{CN} (0, \mathbf{I_m}) $ 
 and
$ \beta_{mljk} $
models path loss and shadowing effects.

 Because large scale fading coefficients changes very slowly and the distance between the antennas of each user is small, so we can  assume that:  
$ \beta_{mljk} = \beta_{lk} $
and for normalization of average power in BS, consider 
$ \beta_{l1} \ge \beta_{l2} \ge...\beta_{lk} $.

Assuming each user selects space-time block code to transmit signals, the received signal at the $ l $-th BS will be as follows:
\begin{equation}
\mathbf{Y}_l = \sum_{k=1}^{K}(\sqrt{\frac{\rho}{2}} \mathbf{H}_{lk} \mathbf{B}_{lk} \mathbf{X}_{lk}) + \mathbf{W}_l
\label{f1},
\end{equation}
where 
$ \mathbf{H}_{lk} = (h_{mljk})_{M \times 2} $
and
$ \mathbf{B}_{lk} = \beta_{lk} \mathbf{I}_2 $.
$ \rho $ shows the signal to noise ratio (SNR) of received signal and 
$ \frac{1}{\sqrt{2}} $
is the normalization factor of the transmitted signal energy  at each time slot. 
$ [\mathbf{X}]_{2 \times T} $
,
$ [\mathbf{W}]_{M \times T} $
and
$ [\mathbf{Y}]_{M \times T} $
are the $ k $-th user transmitted-code, noise and received signal matrices, respectively, where $ M $ represents the number of each BS antennas and $ T $ is the number of time slots. Noise matix entries are i.i.d., which have a complex Gaussian distribution with zero-mean and unit variance.

\section{GOLDEN CODE, LINEAR DECODERS AND NEUMANN SERIES}
\label{ff}
In this section, we first discuss the concepts of code design and linear decoders. In the next subsection, we introduce the method of computational complexity reduction using the Neumann series.
\subsection{GOLDEN CODE AND LINEAR DECODERS}
The Golden code was first introduced in 
\cite{m9}
as a space-time block code for a 
$ 2 \times 2 $
MIMO system. This code is made by a special set of algebra called cyclic division algebra. The transmitted  signal matrix  using the Golden code in the $ T=2 $ time slot is as follows:
\begin{equation}
\mathbf{X}_k = 
\begin{bmatrix}
a_k(x_{k1} + b_k x_{k2}) & \gamma_k a_k(x_{k3} + b_k x_{k4} ) \\
c_k(x_{k3} + d_k x_{k4}) & c_k(x_{k1} + d_k x_{k2})
\end{bmatrix}
\label{f2},
\end{equation}
Where
$ x_{kl} $
coefficients are the distinctive symbols that $ k $-th user transmits and 
$ a_k, b_k, c_k, d_k, \gamma_k $
are constants.

One of the important properties of large number of algebraic structures of space-time codes  is the exchangeability of structure (correlations) that is expressed as, if the system model matrices are
$ \mathbf{X} \in C^{M \times T} $
,
$ \mathbf{H} \in C^{N \times M} $
,
$ \mathbf{Y} \in C^{N \times T} $
and
$ \mathbf{W} \in C^{N \times T} $, 
They can be shown as follows:
$ \mathbf{x} \in C^{L \times 1} $
,
$ \mathbf{\tilde{H}} \in C^{NT \times L} $
,
$ \mathbf{\tilde{y}} \in C^{NT \times 1} $
and
$ \mathbf{\tilde{w}} \in C^{NT \times 1} $, 
Where 
$ \mathbf{\tilde{H}} $
contains both the channel coefficients and the effect of transmitted code. $ L $ is the number of distinct transmitted symbols in each code block in $ T $ time slots. The exchangeability of structure is feasible for linear dispersion codes with real-valued dispersion matrices and Golden code
\cite{m14}.

Considering the $ k $-th user matrix as 
$ \mathbf{H}_{lk} = [\mathbf{h}_{lk1}  , \mathbf{h}_{lk2}] $
and with the 
\eqref{f2}
insertion into 
\eqref{f1}
and the Golden code exchangeability of structure feature, we will have the following relation for the received signal: 
\begin{equation}
Vec(\mathbf{Y}_{l}) = \sum_{k=1}^{K} \sqrt{\frac{\rho}{2}} \beta_{lk} \mathbf{\tilde{H}}_{lk} \mathbf{x}_k  + Vec(\mathbf{W}_l)
\label{f3},
\end{equation}
Where the $ k $-th user channel matrix is as follows:
\begin{equation}
\mathbf{\tilde{H}}_{lk } = 
\begin{pmatrix}
a_k \mathbf{h}_{lk1} & a_k b_k \mathbf{h}_{lk1} & c_k \mathbf{h}_{lk2} & c_k d_k \mathbf{h}_{lk2} \\
c_k \mathbf{h}_{lk2} & c_k d_k \mathbf{h}_{lk2} & \gamma_k a_k \mathbf{h}_{lk1} & \gamma_k a_k b_k \mathbf{h}_{lk1}
\end{pmatrix}
\label{f4},
\end{equation}
and
$ \mathbf{x}_k = (x_{k1}, x_{k2}, x_{k3}, x_{k4})^t $
is the $ k $-th user vector.

The used Golden code coefficients, according to   
\cite{m8}, \\
are as follow:
\begin{equation}
\begin{split}
a_k = (1 + j(1 - b_k))/\sqrt{5}\\
b_k = (1 + \sqrt{5})/2\\
c_k = (1 + j(1 - d_k))/\sqrt{5} \\
d_k = (1 - \sqrt{5})/2 \\
\gamma_k = j.
\end{split}
\end{equation}
According to large-$ M $ analysis ($ M  \to \infty$) , it is easily proven that:
\begin{equation}
\begin{split}
& \frac{\mathbf{\tilde{H}}_{lk }^H \mathbf{\tilde{H}}_{lk }}{M} \xrightarrow{a.s} diag(p,s,p,s),  \\
& \frac{\mathbf{\tilde{H}}_{lk }^H \mathbf{\tilde{H}}_{lk^{\prime} }}{M} \xrightarrow{a.s} \mathbf{0}, 
\end{split}
\end{equation}
where 
$ p=|a_k|^2 + |c_k|^2 $,
$ s = |a_k b_k|^2 + |c_k d_k|^2 $
 and "a.s"  represents "almost surely"
\cite{m8}.

We define the matrices of the system coefficients as follows:
\begin{equation}
\begin{split}
\mathbf{\tilde{G}}_l &= [ \beta_{l1} \mathbf{\tilde{H}}_{l1}, \dots ,\beta_{lk} \mathbf{\tilde{H}}_{lk}], \\
\mathbf{\tilde{X}}_l &= (\mathbf{x^t}_{l1} , \mathbf{x^t}_{l2} , \dots , \mathbf{x^t}_{lk})^t,
\end{split}
\label{f5}
\end{equation}
where 
$ \mathbf{\tilde{G}} $
and
$ \mathbf{\tilde{x}} $
are the
$ 2M \times 4K $
matrix and 
$ 4K $
vector, respectively.

Then we can write equation 
\eqref{f3}
 as follows:
\begin{equation}
Vec(\mathbf{Y}_l) = (\sqrt{\frac{\rho}{2}} \mathbf{\tilde{G}}_l  \mathbf{\tilde{x}}_l) + Vec(\mathbf{W}_l)
\label{f6}.
\end{equation}
According to the decoder matrices, the decoding by the 
$ \mathbf{A}_l $
matrix is as follows:
\begin{equation}
\mathbf{A}_l Vec(\mathbf{Y}_l) = (\sqrt{\frac{\rho}{2}} \mathbf{A}_l \mathbf{\tilde{G}}_l  \mathbf{\tilde{x}}_l) +  \mathbf{A}_l Vec(\mathbf{W}_l)
\label{f7},
\end{equation}
Where A matrix is
\begin{equation}
\mathbf{A}_l = \left \{
\begin{array}{lr}
\mathbf{(\tilde{G}}^H_l \mathbf{\tilde{G}})^{-1}_l \mathbf{\tilde{G}}^H_l  &  \mathbf{ZF}, \\	  
(\mathbf{\tilde{G}}^H_l \mathbf{\tilde{G}}_l + 2 \mathbf{I}_{4K}/\rho)^{-1} \mathbf{\tilde{G}}^H_l &  \mathbf{MMSE}.
\end{array} \right. 
\label{f8}
\end{equation}

\subsection{ NEUMANN SERIES}
Define the 
$ \mathbf{Z}_l $
matrix as follow:
\begin{equation}
\mathbf{Z}_l = \left \{
\begin{array}{lr}
\mathbf{\tilde{G}}^H_l \mathbf{\tilde{G}}_l   &  \mathbf{ZF}, \\

\mathbf{\tilde{G}}^H_l \mathbf{\tilde{G}}_l + 2 \mathbf{I}_{4K}/\rho &  \mathbf{MMSE}.
\end{array} \right. 
\label{f9}
\end{equation}
Much of the computational complexity of linear decoding is on the inverse Hermitian matrix  
$ \mathbf{Z}_l \in \mathbb{C}^{2M \times 2M} $
calculation.
It is also worth noting that the computation of 
$ \mathbf{Z}^{-1}_l $
by usual methods such as Cholesky decomposition needs  
$ \mathcal{O} (M^3)  $
computation,
therefore implementation of such algorithms to serve many users may have some limitations. In terms of hardware constraints, an algorithm based on the Neumann series was first proposed in
\cite{m17} 
to approximate the
$ \mathbf{Z}_l $ matrix .
According to 
\cite{m15}
 if 
$ \mathbf{Z}_l \mathbf{= D + E} $
where 
$ \mathbf{D} $
is a diagonal matrix with the diagonal entries of 
$ \mathbf{Z}_l $
and
$ \mathbf{E} $
is the remainder of the 
$ \mathbf{Z}_l $ 
matrix entries 
, then the Neumann series for inverse calculation will be as follows:
\begin{equation}
\mathbf{\tilde{Z}}^{-1}_{lR} = \sum_{r=0}^{R-1} \mathbf{(-D^{-1} E)^r D^{-1}}
\label{f10},
\end{equation}
where $ R $ is the order of the Neumann series
and 
$ \mathbf{\tilde{Z}}^{-1}_{lR} $
is the $ R $-term approximation of 
$ \mathbf{Z}^{-1}_l $.
 If the maximum modulus of eigenvalues of matrix
$ (\mathbf{I} - \mathbf{D}^{-1} \mathbf{Z}_l) $
is less than 1, then \eqref{f10} converges,
 and also  the approximation will be closer to 
$ R \to \infty $
\cite{m15}.
Moreover, convergence will occur faster if the eigenvalues are lower,
That would be true as long as the ratio 
$ \gamma = \frac{M}{K} $
is high
\cite{m16}.
Neumann series is a iterative method with low computational complexity, So unlike the usual methods 
it is hardware friendly
\cite{m15}.
For example, the approximation for $ R=2 $ will be as follows:
\begin{equation}
\mathbf{\tilde{Z}}^{-1}_{l2} = \underbrace{\mathbf{D}}_{p_0} - \underbrace{(\mathbf{D}^{-1} \mathbf{E}) \mathbf{D}^{-1}}_{p_1}
\label{f19}
\end{equation}
The number of calculations is related to the $ p_0 $ part is $ 2M $ divisions and for part $ p_1 $ is $ 12M^2 - 6M $ Multiplications, while calculating $ \mathbf{Z}_l $ with exact methods requires $ \mathcal{O} (M^3) $ computation.

\section{SPECTRAL EFFICIENCY AND BER OF THE SYSTEM}
\label{fff}
In this section, we investigate the system spectral efficiency and BER, when dual-antenna users exploying Golden code exchange information with the BSs and  BSs, after decoding, sends the information to the CPU for final processing.

\subsection{BER PERFORMANCE}
Estimated signal 
$ x_{lki} $
will be as follows: 
\begin{equation}
\begin{split}
\hat{x}_{lki} =& argmin_x   \\
&  \left \| (\mathbf{A}_l vec(\mathbf{Y}_{4(k-1) + i}))_l - \sqrt{\frac{\rho}{2}} x_l (\tilde{a}_{4(k-1) + i,4(k-1) + i})_l  \right \|,
\label{f11}
\end{split}
\end{equation}
where
$ \tilde{a}_{i,j} = (\mathbf{A}_l \mathbf{\tilde{G}}_l)_{i,j}  $
and
$ x_l $
 represents the $ l $-th entry of vector
$ \mathbf{x} $
.

Each BS quantizes the user data and forwards quantized signals to the CPU to
extract the final data of each user.
\subsection{ SPECTRAL EFFICIENCY}
Using equations 
\eqref{f6}
and 
\eqref{f11}
we have  the following equation for the arbitrary user.
\begin{equation}
s_{k(i),l} = (\mathbf{A} Vec(\mathbf{Y}_{k(i)}))_l = \underbrace{\sqrt{\frac{\rho}{2}}   (\tilde{a}_{k(i),k(i)} x_{k(i)})_l}_{Desired Signal}  +  
\underbrace{({\mathbf{\tilde{w}}}_{k(i)})_l}_{N/I}
\label{f12},
\end{equation}
where
\begin{equation}
{\mathbf{\tilde{w}}}_{k(i),l} =  \sum_{\substack{t=1 \\ t \neq k(i)}}^{4K} \underbrace{ \sqrt{\frac{\rho}{2}} (\tilde{a}_{k(i),t})_l (x_t)_l}_{Interference} + \underbrace{((\mathbf{A} Vec(\mathbf{W}))_{k(i)})_l}_{Noise}
\label{f13},
\end{equation}
where 
$ k(i) $ 
is the $ i $-th symbol of $ k $-th user and it is expressed as
$ k(i) = 4(k - 1) + i $.
As discussed in 
\ref{f}
, the received data at each BS are independently decoded and then these signals are transmitted to the CPU via a back-haul link. As a result, the signal required to extract $ k $-th user data is equal to 
$ r_{k(i)} $, 
that is expressed as follows:
\begin{equation}
r_{k(i)} = \sum_{l=1}^{L} s_{k(i),l}
\label{f14}.
\end{equation}

To calculate the spectral efficiency of the whole system according to the above relation, we need to calculate the variance of the noise and interference. Due to the independence of the transmitted signals with each other,  
$ \mathbf{E}(|x_t|^2)=1 $
and their independence from 
$ \mathbf{W} $
, according to
\eqref{f12} and \eqref{f14},
we can write as follow:
\begin{equation}
\mathbf{E}(|\mathbf{\tilde{w}}_{k(i)}|^2) = \frac{\rho}{2} \sum_{l=1}^{L} \sum_{\substack{t=1 \\ t \neq k(i)}}^{4K} (|(\mathbf{a}_{k(i),t} \tilde{\mathbf{g}}_{t})_l|^2 + |(\mathbf{a}_{k(i)})_l|^2)
\label{f15},
\end{equation}
where
$ \mathbf{a}_i $
and
$ \tilde{\mathbf{g}}_j $
are the $ i $-th row elements of matrix
$ \mathbf{A}_l $
and $ j $-th column elements of matrix 
$ \mathbf{G}_l $
, respectively.

The Signal to Interference and Noise Ratio(SINR) will be calculated as follows:
\begin{equation}
SINR_{k(i)} = \frac{\frac{\rho}{2} \sum_{l=1}^{L} |(\mathbf{a}_{k(i),k(i)} \tilde{\mathbf{g}}_{k(i)})_l |^2 }{\frac{\rho}{2} \sum_{l=1}^{L} \sum_{\substack{t=1 \\ t \neq k(i)}}^{4K} (|(\mathbf{a}_{k(i),t} \tilde{\mathbf{g}}_t)_l|^2 + |(\mathbf{a}_{k(i)})_l|^2)}
\label{f16}.
\end{equation}
The spectral efficiency  lower bound in uplink for an arbitrary user is as follows:
\begin{equation}
SE_k \ge \frac{1}{2} \sum_{i=1}^{4} log_2 (1 +SINR_{k(i)} )
\label{f17},
\end{equation}
where factor
$ \frac{1}{2} $
comes from the fact that we send four symbols in two time slots.

\section{SIMULATION RESULTS}
\label{ffff}
In this section, we analyze our theoretical results with simulation. In this simulations users with golden code exchange information with BSs . We consider the number of BSs to be four and the users can connect to all BSs.

\subsection{BER SIMULATION}
In this simulation, the BER of the proposed system with the usual (exact) structure and the approximate calculation of the inverse matrix using Neumann series with 
$ R=2 $
is investigated. Assume that  
$ M=256 $
 and 
$ K=10 $.
 Each user has two antennas. The used modulation for each user is BPSK. The ZF and MMSE are the decoders used at each BS. Results are compared  with the state which users  have only single antenna. The used modulation for any user in this mode is 4QAM  for providing  fairness in bandwidth efficiency compared to the dual-antenna mode.
\begin{figure}[tb]
	\centerline{\includegraphics[scale=0.24]{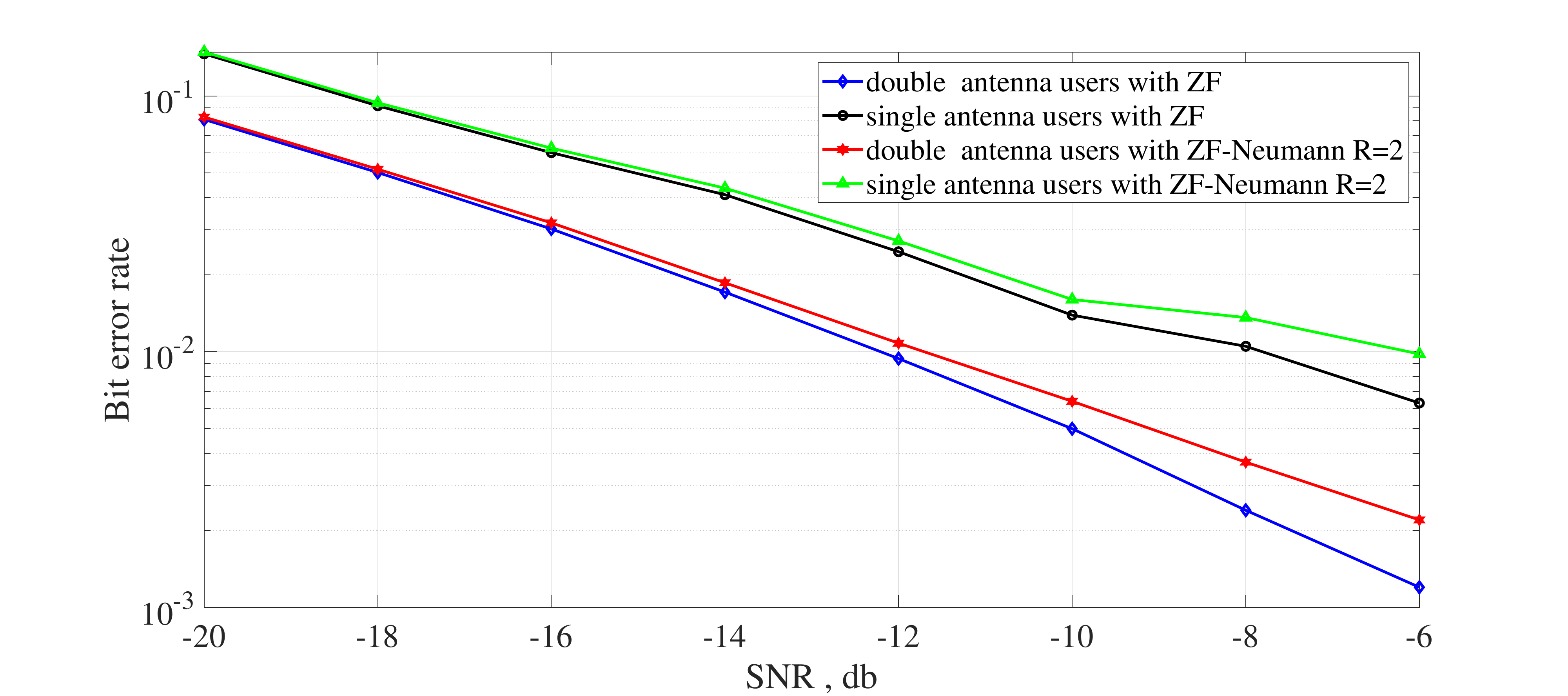}}
	\caption{The system BER  with the ZF and ZF-Neumann series with R=2 decoder. $ M=256 $ and  $ K=10$}
	\label{fig2}
\end{figure}
\begin{figure}[tb]
	\centerline{\includegraphics[scale=0.24]{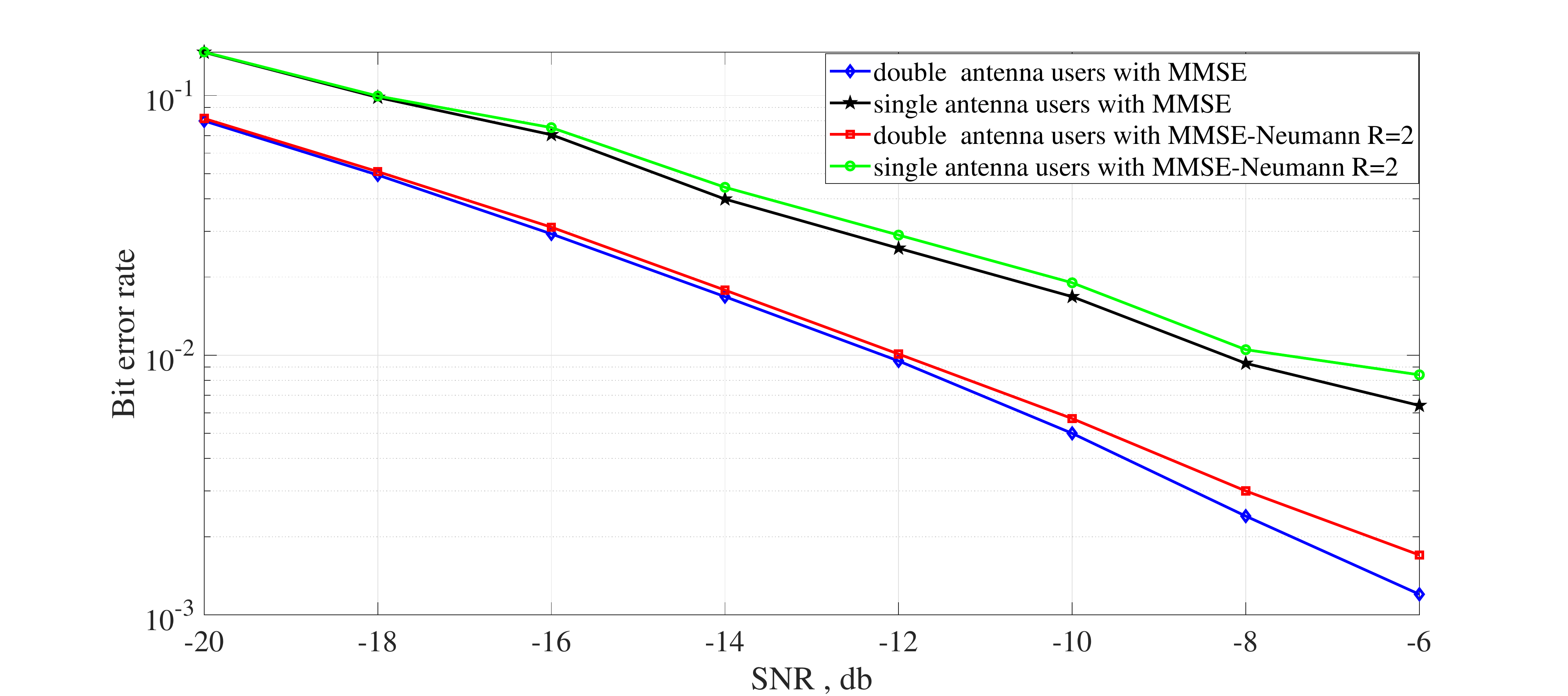}}
	\caption{The system BER  with the MMSE and MMSE-Neumann series with R=2 decoder. $ M=256 $ and  $ K=10$}
	\label{fig3}
\end{figure}
The change of BER versus SINR are described in Fig.
\ref{fig2}
and  Fig. 
\ref{fig3}. 
As can be seen from the figures, the system BER with dual-antenna users have better performance than a single-antenna mode. This indicates that  diversity gain in uplink is earned for dual-antenna users, and it is seen in the figures that the approximation method using the Neumann series with 
$ R=2 $
based on \eqref{f19},
performs close to the exact structure. 

\subsection{SPECTRAL EFFICIENCY SIMULATON }
\begin{figure}[tb]
	\centerline{\includegraphics[scale=0.24]{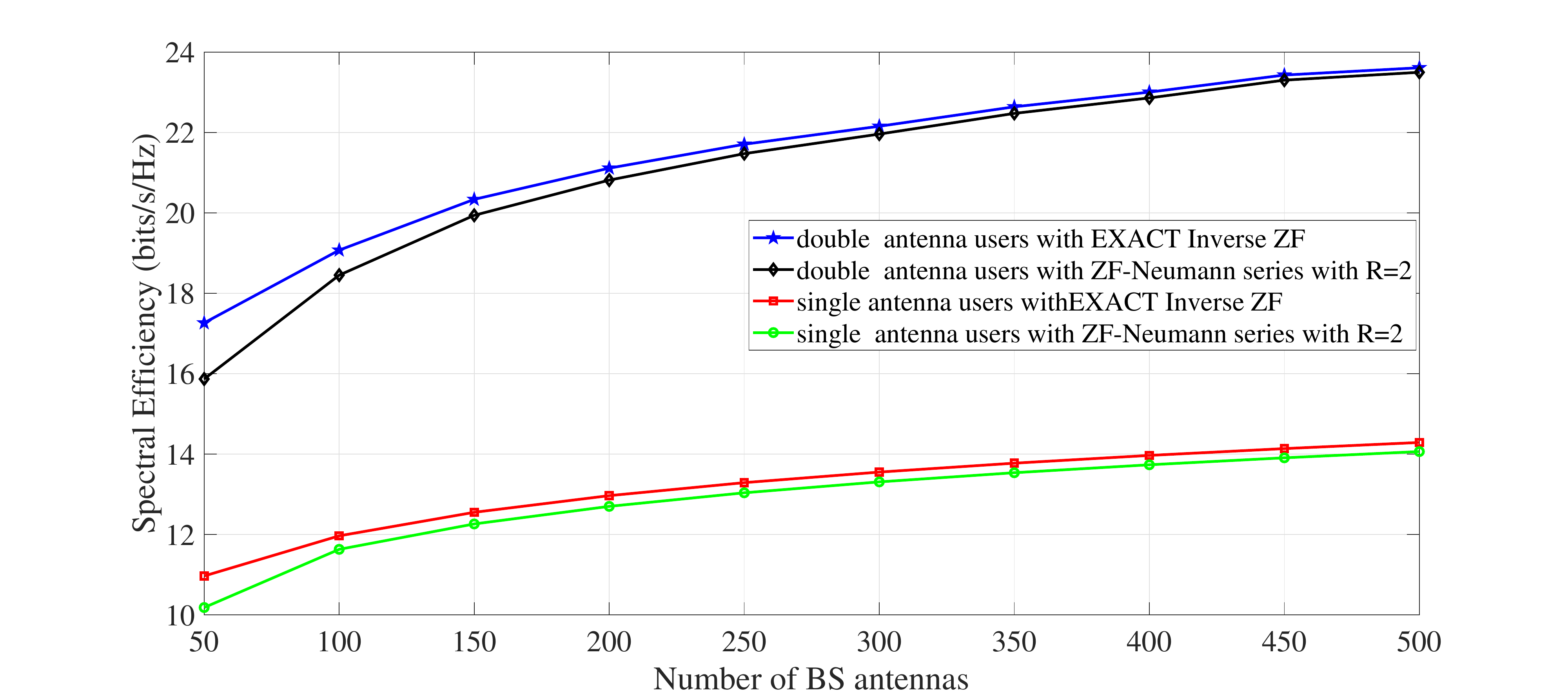}}
	\caption{The system spectral efficiency  with ZF decoder.  $ K=10$}
	\label{fig4}
\end{figure}

In this simulation, we investigate the spectral efficiency of the system with the linear decoders.
Assume that
$ K=10 $ 
 and $\rho = 10 $. Large scale fading coefficients $\beta_{lk}$ are chosen uniformly and randomly  in interval 
$ [0, 1] $.
  BS antennas varies from 50 to 500. The simulation is based on 
\eqref{f16}
and
\eqref{f17} formulas.
The change of spectral efficiency versus Bs antennas are described in Fig. 
\ref{fig4}.
We have also performed a single-antenna users mode based on 
$ (9) $ and $ (27) $
in paper
\cite{m18}
to compare performance.
As can be seen from the Fig.
\ref{fig4}
, dual-antenna users have better performance in term of spectral efficiency  than single-antenna users, and  the approximation method using the Neumann series with 
$ R=2 $
based on \eqref{f19},
performs close to the exact method, that is a completely hardware-friendly approach to reduce computational complexity of the decoder.

\section{CONCLUSION}
\label{dd}
In this paper the performance of multiuser cell-free massive MIMO system with dual-antenna users using space-time block codes in terms of BER and spectral efficiency with linear decoders, ZF and MMSE were studied. The lower bound of throughput for the asummed system was derived. The simulation results show that the dual-antenna mode performs better than single-antenna mode in terms of BER and spectral efficiency in the same system. Also we showed that the performance of a given system with the proposed method with less computational complexity using neumann series with 
$ R=2 $
was close to the exact matrix inverse.
\bibliographystyle{ieeetr}
\bibliography{documentref}
\balance

\end{document}